\documentclass[letterpaper,pra,superscriptaddress,floatfix,pra,amsmath,amsfonts,amssymb,%
twocolumn,showpacs,tightenlines]{revtex4}

\usepackage{graphicx}
\usepackage{bm}
\usepackage{empheq}

\def\1{\mathit 1}

\def\intzT{\int_0^T}
\def\infint{\int_{-\infty}^{\infty}}
\def\halfint{\int_0^\infty}

\def\bra#1{\left\langle {#1} \right\rvert}
\def\ket#1{\left\lvert {#1} \right\rangle}

\def\outprod#1#2{\ket {#1}\!\bra {#2}}

\def\avg#1{\left\langle {#1} \right\rangle}
\def\abs#1{\left\lvert{#1}\right\rvert}

\def\c#1{\mathcal{#1}}

\begin{document}

\title{A Single Trapped Ion as a Time-Dependent Harmonic Oscillator}
 
\author{Nicolas~C.~Menicucci}
\email{nmen@princeton.edu}
\affiliation{Department of Physics, Princeton University, Princeton, NJ 08544, USA}
\affiliation{School of Physical Sciences, The University of Queensland, Brisbane, Queensland 4072, Australia}

\author{G.~J.~Milburn}
\affiliation{School of Physical Sciences, The University of Queensland, Brisbane, Queensland 4072, Australia}

\date\today

\begin{abstract}
We show how a single trapped ion may be used to test a variety of important physical models realized as time-dependent harmonic oscillators.  The ion itself functions as its own motional detector through laser-induced electronic transitions.  Alsing {\it et al.\/}\ [Phys.\ Rev.\ Lett.\ {\bf 94}, 220401 (2005)] proposed that an exponentially decaying trap frequency could be used to simulate (thermal) Gibbons-Hawking radiation in an expanding universe, but the Hamiltonian used was incorrect.  We apply our general solution to this experimental proposal, correcting the result for a single ion and showing that while the actual spectrum is different from the Gibbons-Hawking case, it nevertheless shares an important experimental signature with this result.
\end{abstract}


\maketitle


\section{Introduction}
\label{sec:intro}

The time-dependent quantum harmonic oscillator has long served as a paradigm for nonadiabatic time-dependent Hamiltonian systems and has been applied to a wide range of physical problems by choosing the mass, the frequency, or both, to be time-dependent. The earliest application is to squeezed state generation in quantum optics~\cite{Stoler1970,Yuen1976,Hollenhorst1979}, in which the effect of a second-order optical nonlinearity on a single-mode field can be modeled by a harmonic oscillator with a frequency that is harmonically modulated at twice the bare oscillator frequency. It was subsequently shown that any modulation of the frequency could produce squeezing~\cite{Ma1989}, and thus the same model could be used to approximately describe the generation of photons in a cavity with a time-dependent boundary~\cite{Moore1970,Dodonov1996}. 

The model has been used in a number of quantum cosmological models. In Ref.~\cite{Kim1995}, a time-dependent frequency has been used to explain entropy production in a quantum mini-superspace model. The model, with both mass and frequency time-dependent, has been particularly important in developing an understanding of how quantum fluctuations in a scalar field can drive classical metric fluctuation during inflation~\cite{Polarski1996,Kiefer1998}. In a cosmological setting the time-dependence is not harmonic and is usually exponential.  In all physical applications, of course, the model is only an approximation to the true physics, and its validity can be tested only with considerable difficulty, especially in the cosmological setting. Here we propose a realistic experimental context in which the time-dependent quantum harmonic oscillator can be studied directly. 

Many decades of effort to refine spectroscopic measurements for time standards now enable a single ion to be confined in three dimensions, its vibrational motion restricted effectively to one dimension, and the ion cooled to the vibrational ground state with a probability greater than 99\%~\cite{Leibfried2003}.  Laser cooling is based on the ability to couple an internal electronic transition to the vibrational motion of the ion~\cite{Monroe1995}. These methods can easily be extended to more than one ion and their collective normal modes of vibration~\cite{James1998}. Indeed so carefully can the coupling between the electronic and vibrational states be engineered that is is possible to realise simple quantum information processing tasks~\cite{Leibfried2003a,Schmidt-Kaler2003}.  We use the control of trapping potential afforded by ion traps, together with the ability to reach quantum limited motion, to propose a simple experimental test of quantum harmonic oscillators with time-dependent frequencies. We also make use of the ability to make highly efficient quantum measurements, based on fluorescent shelving~\cite{Leibfried2003}, to propose a practical means to test our predictions.
 
In this paper, we calculate the excitation probability of a trapped ion in a general time-dependent potential.  When beginning in the vibrational ground state of the unchirped trap and starting the chirping process adiabatically, the excitation probability is simply related to the Fourier transform of the solution of the Heisenberg equations of motion (which is also the same as the trajectory of the equivalent classical oscillator).  We compare our result with that of Ref.~\cite{Alsing2005} for the case of a single ion undergoing an exponential frequency chirp.  The cited work attempts to use this experimental setup to model a massless scalar field during an inflating (i.e., de~Sitter) universe, which would give a thermal excitation spectrum as a function of the detector response frequency~\cite{Gibbons1977}.  The analysis is incorrect, however, because the wrong Hamiltonian was used.  Nevertheless, the corrected calculation presented here also gives an excitation spectrum with a thermal signature, although the particular functional form is different.


\section{General Solution}
\label{sec:gensol}

The quantum Hamiltonian for a single ion in a time-dependent harmonic trap can be well-approximated in one dimension by
\begin{align}
\label{eq:H}
	H = \frac {p^2} {2M} + \frac M 2 \nu(t)^2 q^2\;,
\end{align}
where $\nu(t)$ is time-dependent but always assumed to be much slower than the timescale of the micromotion~\cite{Leibfried2003}.  For emphasis, we have indicated the explicit time-dependence of the frequency~$\nu$; we will often omit this from now on.  Working in the Heisenberg picture, we get the following equations of motion for~$q$ and~$p$:
\begin{align}
\label{eq:qdot}
	\dot q &= \frac p M\;, \\
\label{eq:pdot}
	\dot p &= -M \nu^2 q\;.
\end{align}
Dots indicate total derivatives with respect to time.  Differentiating again and plugging in these results gives
\begin{align}
\label{eq:qdoubledot}
	0 &= \ddot q + \nu^2 q\;, \\
\label{eq:pdoubledot}
	0 &= \ddot p - 2 \frac {\dot \nu} {\nu} \dot p + \nu^2 p\;.
\end{align}
As we shall see, only Eq.~\eqref{eq:qdoubledot} is necessary for calculating excitation probabilities, so we will focus only on it.  These equations are operator equations, but they are identical to the classical equations of motion for the analogous classical system.  Interpreting them as such, we will label the two linearly independent c-number solutions as $h(t)$ and $g(t)$, where the following initial conditions are satisfied:
\begin{align}
\label{eq:ICs}
	h(0) = \dot g(0) = 1 \qquad \text{and} \qquad \dot h(0) = g(0) = 0\;,
\end{align}
Writing $q(0) = q_0$ and $p(0) = p_0$, the unique solution for $q$ to the initial value problem above is
\begin{align}
\label{eq:qeom}
	q(t) = q_0 h(t) + \frac {p_0} {M} g(t)\;.
\end{align}
By differentiating and using the relations above, we know also that
\begin{align}
\label{eq:peom}
	p(t) = M q_0 \dot h(t) + p_0 \dot g(t)\;.
\end{align}
To check our math, we can verify that $[q(t),p(t)] = i\hbar$, which is fulfilled if and only if the Wronskian $W(h,g)$ of the two solutions is one for all times---specifically,
\begin{align}
\label{eq:hgcommutator}
	h \dot g - \dot h g = 1\;,
\end{align}
where we have assumed that $[q_0, p_0] = i\hbar$.

Moreover, if the initial state at $t=0$ is symmetric with respect to phase-space rotations, then we have additional rotational freedom in choosing the initial quadratures.  (This would be the case, for instance, if we start in the instantaneous ground state.)  Notice that Eq.~\eqref{eq:qeom} can be written as the inner product of two vectors:
\begin{align}
\label{eq:qinprod}
	q(t) = \Bigl( q_0, \frac {p_0} {M \nu_0} \Bigr) \cdot \Bigl( h(t), \nu_0 g(t) \Bigr)
\end{align}
(and similarly for Eq.~\eqref{eq:peom}), where we have normalized the quadrature operators to have the same units.  As an inner product, this expression is invariant under simultaneous rotations of both vectors.  Thus, if the initial state possesses rotational symmetry in the phase plane, then the rotated quadratures are equally as valid as the original ones for representing the initial state, which means that an arbitrary rotation can be applied to the second vector above without changing any measurable property of the system.  This freedom can be used, for instance, to define new functions~$h'(t)$ and~$g'(t)$ that are more convenient for calculations, where the linear transformation between them and the original ones (with prefactors as in Eq.~\eqref{eq:qinprod}) is a rotation.  We will use this freedom in the next section.

One reason why ion traps have become a leading implementation for quantum information processing is the ability to efficiently read out the internal electronic state using a fluorescence shelving scheme~\cite{Leibfried2003}. As the internal state can become correlated with the vibrational motion of the ion, this scheme can be configured as a way to measure the vibrational state directly~\cite{Wallentowitz1996}. To correlate the internal electronic state with the motion of the ion, an external laser can be used to drive an electronic transition between two levels~$\ket g$ and~$\ket e$, separated in energy by $\hbar\omega_A$. The interaction between an external classical laser field and the ion is described, in the dipole and rotating-wave approximation, by the interaction-picture Hamiltonian~\cite{Leibfried2003}
\begin{align}
\label{eq:HLdef}
	H_L = -i\hbar\Omega_0 \left[\sigma_+(t)e^{ik\cos\theta q(t)}-\sigma_-(t)e^{-ik\cos\theta q(t)}\right]\;,
\end{align}
where $\Omega_0$ is the Rabi frequency for the laser-atom interaction, $\omega_L$ is the laser frequency, $k$ is the magnitude of the wave vector $\vec k$, which makes an angle $\theta$ with the trap axis, $q(t)$ is given in Eq.~\eqref{eq:qeom}, and
\begin{equation}
\label{eq:sigmapmdef}
	\sigma_\pm(t) = e^{\pm i \Delta t} \sigma_\pm\;.
\end{equation}
The electronic-state raising and lowering operators are defined as $\sigma_+=\outprod e g$ and $\sigma_-=\outprod g e$, respectively, and
\begin{equation}
\label{eq:Deltadef}
	\Delta=\omega_A-\omega_L
\end{equation}
is the detuning of the laser below the atomic transition.  We can construct a meaningful quantity that characterizes the ``size'' of $q(t)$ based on the width of the ground-state wave packet for an oscillator with frequency $\nu(t)$, namely $\sqrt{\hbar/2 M \nu(t)}$.  As long as this quantity is much smaller than $k \cos \theta$ throughout the chirping process, then we can expand the exponentials in Eq.~\eqref{eq:HLdef} to first order and define the interaction Hamiltonian~$H_I$ between the electronic states and vibrational motion (still in the interaction picture) by
\begin{align}
\label{eq:HI}
	H_I = \hbar \Omega_0 k \cos \theta q(t) \bigl( e^{-i \Delta t} \sigma_- + e^{+i \Delta t} \sigma_+ \bigr)\;.
\end{align}
where we have assumed that $\omega_L$ is far off-resonance, and thus $\Delta \not \simeq 0$.

Using first-order time-dependent perturbation theory, the probability to find the ion in the excited state is 
\begin{multline}
\label{eq:Pexcited}
	P^{(1)} = \frac 1 {\hbar^2} \intzT dt_1 \intzT dt_2 \avg { H_I(t_1) \mathcal P_e H_I(t_2) } \\
	= \Omega_0^2 k^2 \cos^2 \theta \intzT dt_1 \intzT dt_2\, e^{-i\Delta (t_1 - t_2)} \avg { q(t_1) q(t_2) }\;,
\end{multline}
where $\mathcal P_e = 1_{\text{vib}} \otimes \outprod e e$ is the projector onto the excited electronic state (and the identity on the vibrational subspace).  We always assume that the ion begins in the electronic ground state.  If the ion also starts out in the instantaneous vibrational ground state for a static trap of frequency $\nu_0 = \nu(0)$ at $t=0$ (which is most useful when the chirping begins in the adiabatic regime), then we can evaluate the two-time correlation function as
\begin{align}
\label{eq:twopointvac}
	&\avg{ q(t_1) q(t_2) }_{\text{ground}} = \avg { q_0^2 } h(t_1) h(t_2) + \frac {\avg { p_0^2 }} {M^2} g(t_1) g(t_2) \nonumber \\
	&\qquad\qquad + \frac {\avg { q_0 p_0 }} {M} \bigl [h(t_1) g(t_2) - h(t_2) g(t_1) \bigr] \nonumber \\
	&\qquad = \frac {\hbar} {2M \nu_0} \Bigl[ h(t_1) - i \nu_0 g(t_1) \Bigr] \Bigl[ h(t_2) + i \nu_0 g(t_2) \Bigr] \nonumber \\
	&\qquad = \frac {\hbar} {2M \nu_0} f(t_1) f^*(t_2)\;,
\end{align}
where we have used the facts that for the vibrational ground state, $\avg{ q_0^2 } = \avg{ (p_0/M\nu_0)^2} = \hbar/2M\nu_0$ and $\avg{ q_0 p_0 } = \tfrac 1 2 \avg{ \{q_0,p_0\} + [q_0, p_0] } = i\hbar /2$, and we have defined the complex function
\begin{align}
\label{eq:fdef}
	f(t) &= h(t) -i \nu_0 g(t)\;,
\end{align}
which is the solution to Eq.~\eqref{eq:qdoubledot} with initial the conditions, $f(0) = 1$ and $\dot f(0) = -i\nu_0$.  Plugging this into Eq.~\eqref{eq:Pexcited} gives, quite simply,
\begin{align}
\boxed{
\label{eq:Pexcitedvac}
	P^{(1)} \to (\Omega_0 \eta_0)^2 \abs{\c F}^2\;,
}
\end{align}
where
\begin{align}
\label{eq:cFdef}
	\c F &= \intzT dt\, e^{-i \Delta t} f(t)\;,
\end{align}
and we have defined the unitless, time-dependent Lamb-Dicke parameter~\cite{Leibfried2003} as
\begin{align}
\label{eq:LambDicke}
	\eta(t) = \sqrt{ \frac {\hbar k^2 \cos^2 \theta} {2 M \nu(t)} }\;,
\end{align}
and $\eta_0 = \eta(0)$.  Recalling that $f(t)$ can be considered a complex c-number solution to the equations of motion for the equivalent classical Hamiltonian, Eq.~\eqref{eq:Pexcitedvac} shows that the excitation probability is simply related to the Fourier transform of the classical trajectories when beginning in the vibrational ground state.


\section{Exponential Chirping}
\label{sec:expchirp}

Recent work~\cite{Alsing2005} has suggested that an exponentially decaying trap frequency has the same effect on the phonon modes of a string of ions as an expanding (i.e., de~Sitter) spacetime does on a one-dimensional scalar field~\cite{Carvalho2004}.  An inertial detector that responds to such an expanding scalar field would register a thermal bath of particles, called Gibbons-Hawking radiation~\cite{Gibbons1977}.  Ref.~\cite{Alsing2005} suggests that the acoustic analog~\cite{Unruh1981} of this radiation could be seen in an ion trap, causing each ion to be excited with a thermal spectrum with temperature $\hbar \kappa/2\pi k_B$, as a function of the detuning~$\Delta$, where $\kappa$ is the trap-frequency decay rate.  The analysis used an incorrect Hamiltonian that neglected squeezing and source terms that have no analog in the expanding scalar field model but which are present when considering trapped ions in this way, and the results are incorrect.  In this section, we revisit this problem and calculate the excitation probability for a single ion in an exponentially decaying harmonic potential, as a function of the detuning~$\Delta$.

We write the time-dependent frequency as~\footnote{The authors of Ref.~\cite{Alsing2005} consider both signs in the exponential, but we will restrict ourselves to the case that allows us to begin chirping in the adiabatic limit.}
\begin{align}
\label{eq:nuexp}
	\nu(t) = \nu_0 e^{-\kappa t}\;.
\end{align}
This results in
\begin{align}
\label{eq:qdoubledotexp}
	\ddot q + \nu_0^2 e^{-2\kappa t} q = 0\;.
\end{align}
Solutions with initial conditions~\eqref{eq:ICs} are
\begin{align}
\label{eq:hexp}
	h(t) &= \frac {\pi \nu_0} {2 \kappa} \left[ J_1 \left( \frac {\nu_0} \kappa \right) Y_0 \left( \frac \nu \kappa \right) - Y_1 \left( \frac {\nu_0} \kappa \right) J_0 \left( \frac \nu \kappa \right) \right]\;, \\
\label{eq:gexp}
	g(t) &= \frac {\pi} {2 \kappa} \left[ -J_0 \left( \frac {\nu_0} \kappa \right) Y_0 \left( \frac \nu \kappa \right) + Y_0 \left( \frac {\nu_0} \kappa \right) J_0 \left( \frac \nu \kappa \right) \right]\;,
\end{align}
where the time dependence is carried in $\nu = \nu(t)$ from Eq.~\eqref{eq:nuexp}, and $J_n$ and $Y_n$ are Bessel functions.  We could plug these directly into the formulas from the last section, but we will simplify the calculations by considering the limits of slow and long-time frequency decay, represented by
\begin{align}
\label{eq:conditions}
	\nu_0 \gg \kappa \qquad \text{and} \qquad \nu_0 e^{-\kappa T} \ll \kappa\;,
\end{align}
respectively.  This allows us to do several things.  First, it allows us to use the usual ground state of the unchirped trap at frequency $\nu_0$ as a good approximation to the ground state of the expanding trap at $t=0$, since at that time the system is being chirped adiabatically.  This is important because it allows the experiment to begin with a static potential, which is useful for cooling.  Second, it allows us to simplify~$h(t)$ and~$g(t)$ using the phase-space rotation freedom discussed above.  Using asymptotic approximations for the Bessel functions in the coefficients,
\begin{align}
\label{eq:J0Y1far}
	J_0 \left( \frac {\nu_0} \kappa \right) \simeq -Y_1 \left( \frac {\nu_0} \kappa \right) &\simeq \sqrt{ \frac {2 \kappa} {\pi \nu_0} } \cos \left( \frac {\nu_0} \kappa - \frac \pi 4 \right)\;, \\
\label{eq:J1Y0far}
	J_1 \left( \frac {\nu_0} \kappa \right) \simeq \phantom{-}Y_0 \left( \frac {\nu_0} \kappa \right) &\simeq \sqrt{ \frac {2 \kappa} {\pi \nu_0} } \sin \left( \frac {\nu_0} \kappa - \frac \pi 4 \right)\;,
\end{align}
we get
\begin{align}
\label{eq:hexpapprox}
	h(t) &\simeq \sqrt { \frac {\pi \nu_0} {2 \kappa} } \Bigl[ \sin \varphi\, Y_0 \left( \frac \nu \kappa \right) + \cos \varphi\, J_0 \left( \frac \nu \kappa \right) \Bigr]\;, \\
\label{eq:gexpapprox}
	\nu_0 g(t) &\simeq \sqrt { \frac {\pi \nu_0} {2 \kappa} } \Bigl[ -\cos \varphi\, Y_0 \left( \frac \nu \kappa \right) + \sin \varphi\, J_0 \left( \frac \nu \kappa \right) \Bigr]\;.
\end{align}
where $\varphi = \nu_0/\kappa - \pi/4$.  Since we are taking the initial state to be the ground state, which is symmetric with respect to phase-space rotations, we can use the freedom discussed in the previous section to undo the rotation represented by Eqs.~\eqref{eq:hexpapprox} and~\eqref{eq:gexpapprox} and define the simpler functions
\begin{align}
\label{eq:hexpsimple}
	h(t) \to h'(t) &= \sqrt { \frac {\pi \nu_0} {2 \kappa} } Y_0 \left( \frac \nu \kappa \right)\;, \\
\label{eq:gexpsimple}
	g(t) \to g'(t) &= \sqrt { \frac {\pi} {2 \kappa \nu_0} } J_0 \left( \frac \nu \kappa \right)\;.
\end{align}
The primes are unnecessary due to the symmetry of the initial state, so we drop them from now on and plug directly into Eq.~\eqref{eq:fdef}:
\begin{align}
\label{eq:fexp}
	f(t) &= \sqrt { \frac {\pi \nu_0} {2 \kappa} } \left[ Y_0 \left( \frac \nu \kappa \right) -i J_0 \left( \frac \nu \kappa \right) \right] \nonumber \\
	&= -i\sqrt { \frac {\pi \nu_0} {2 \kappa} } H^{(1)}_0 \left( \frac \nu \kappa \right)\;,
\end{align}
where $H^{(1)}_n$ is a Hankel function of the first kind.  The integral in Eq.~\eqref{eq:cFdef} can be evaluated in the limits~\eqref{eq:conditions} using techniques similar to those used in Ref.~\cite{Alsing2005}.  First, define
\begin{align}
\label{eq:dummyvars}
	e^\alpha = \frac \nu \kappa\;,\quad \tau = \alpha - \kappa t\;, \quad u = e^\tau\;, \quad \text{and} \quad x = \Delta/\kappa\;.
\end{align}
The integral in question then becomes (neglecting the prefactor)
\begin{align}
\label{eq:X0int}
	&\intzT dt\, e^{-i \Delta t} H^{(1)}_0 \left( \frac \nu \kappa \right) = \intzT dt\, e^{-i\Delta t} H^{(1)}_0 (e^{\alpha - \kappa t}) \nonumber \\
	&\qquad\qquad= \frac 1 \kappa \int_{\alpha - \kappa T}^\alpha d\tau\, e^{-ix(\alpha - \tau)} H^{(1)}_0 (e^\tau) \nonumber \\
	&\qquad\qquad\to \frac {e^{-ix\alpha}} {\kappa} \infint d\tau\, e^{ix\tau} H^{(1)}_0(e^\tau) \nonumber \\
	&\qquad\qquad= \frac {e^{-ix\alpha}} {\kappa} \halfint du\, u^{ix-1} H^{(1)}_0(u)\;.
\end{align}
Inserting a convergence factor with $x \to x - i\epsilon$, and then taking the limit $\epsilon \to 0^+$, we can use the formula
\begin{align}
\label{eq:H0intformula}
	\halfint du\, u^{ix-1} H^{(1)}_0(u) &= -2^{ix} \frac {\Gamma (ix/2)} {(e^{\pi x} - 1) \Gamma(1-ix/2)}
\end{align}
to evaluate
\begin{align}
\label{eq:cFexp}
	\abs{\c F}^2 &= \frac {\pi \nu_0} {2 \kappa} \frac {1} {\kappa^2} \abs{ \frac {\Gamma (ix/2)} {\Gamma(1-ix/2)} }^2 \frac {1} {(e^{\pi x} - 1)^2} \nonumber \\
	&= \frac {2 \pi \nu_0} {\kappa^3 x^2} \frac {1} {(e^{\pi x} - 1)^2}\;.
\end{align}
When plugging in for the dummy variables~\eqref{eq:dummyvars}, this gives
\begin{align}
\boxed{
\label{eq:Pexcitedexp}
	P^{(1)} = (\Omega_0 \eta_0)^2 \frac {2 \pi \nu_0} {\kappa \Delta^2} \frac {1} {(e^{\pi \Delta/\kappa} - 1)^2}\;.
}
\end{align}
The calculated result from Ref.~\cite{Alsing2005} for a single ion is
\begin{equation}
\label{eq:Pcalcinflation}
	P^{(1)}_{\text{GH}} = (\Omega_0 \eta_0)^2  \frac {2 \pi} {\kappa \Delta} \frac {1} {e^{2\pi\Delta/\kappa}-1} \;,
\end{equation}
which contains a Planck factor with Gibbons-Hawking~\cite{Gibbons1977} temperature $T= \hbar \kappa/2 \pi k_B$ but is different from the actual result for a single ion, given by Eq.~\eqref{eq:Pexcitedexp}.

Several things should be noted about these functions.  First, they both break down as $\Delta \to 0$ because of the approximation made in obtaining Eq.~\eqref{eq:HI}.  They also fail if the time-dependent Lamb-Dicke parameter~\eqref{eq:LambDicke} ever becomes too large throughout the chirping process.  Furthermore, most cases of interest will be $\Delta \simeq \nu_0$ (the first red sideband) and near $\Delta \simeq -\nu_0$ (the first blue sideband), which means that $\abs \Delta \gg \kappa$, since $\nu_0 \gg \kappa$.  The first red sideband represents a detector that requires the absorption of one phonon (plus one laser photon) in order to excite the atom---the usual thing we mean by ``particle detector'' when the particles are phonons.  The first blue sideband, on the other hand, represents a detector that {\em emits} a phonon in order to excite the atom (along with absorbing one laser photon).

There are a couple of ways to compare these functions.  First, we can take the ratio of the two for both the red- and blue-sideband cases.  In both cases, we obtain
\begin{align}
\label{eq:P1P1thermalratio}
	\frac {P^{(1)}} {P^{(1)}_{\text{GH}}} \simeq \frac {\nu_0} {\abs \Delta} (1 + 2e^{-\pi \abs \Delta/\kappa})
\end{align}
plus terms of order $O(e^{-2\pi \abs \Delta/\kappa})$.  Since $\abs \Delta \simeq \nu_0$, the prefactor is close to one, and the second term is very small (since $\nu_0 \gg \kappa$).  Furthermore, it is cumbersome to directly compare the measured probability to the full function (with all the prefactors).  It is often easier instead to make measurements on both the first red sideband and the first blue sideband and then take the ratio of the two.  The constant prefactors disappear in this calculation, and both functions then have the same experimental signature:
\begin{align}
\label{eq:P1redblueratio}
	\frac {P^{(1)}(\Delta)} {P^{(1)}(-\Delta)} = \frac {P^{(1)}_{\text{GH}}(\Delta)} {P^{(1)}_{\text{GH}}(-\Delta)} = e^{-2\pi \Delta/\kappa}\;,
\end{align}
which is that of a thermal distribution with temperature $T = \hbar \kappa/2\pi k_B$, which is of the Gibbons-Hawking form~\cite{Gibbons1977} with the expansion rate given by~$\kappa$.  Therefore, although the Hamiltonian used in the calculations in Ref.~\cite{Alsing2005} was missing terms, the intuition (at least for a single ion) was correct in that the actual experimental signature in this case matches that of an ion undergoing thermal motion in a static trap, where the temperature is proportional to~$\kappa$.

To see whether this experiment is feasible, we must examine the validity of our approximations.  For a typical trap, we expect that $\nu_0 \simeq 1$~MHz, and thus if we take $\kappa \simeq 1$~kHZ, we easily satisfy the first of conditions~\eqref{eq:conditions}, namely $\nu_0 \gg \kappa$. The second of these conditions gives a constraint on the modulation time~$T$. For these parameters we expect that $T \simeq $ a~few~msec. This is compatible with typical cooling and readout time scales and is less than those for heating due to fluctuating patch potentials~\cite{Leibfried2003}.  Thus, this is a realizable experiment with current technology.


\section{Conclusion}
\label{sec:conclusion}

We have shown that a single trapped ion in a modulated trapping potential can serve as an experimentally accessible implementation of a quantum harmonic oscillator with time-dependent frequency, including robust control over state preparation, manipulation, and measurement.  The ion itself serves both as the oscillating particle and as the local detector of vibrational motion via coupling to internal electronic states by an external laser.  For the case of a general time-dependent trap frequency, we calculated the first-order excitation probability for the ion in terms of the solution to the classical equations of motion for the equivalent classical oscillator.  We applied this general result to the case of exponential chirping and corrected the calculation in Ref.~\cite{Alsing2005} for a single ion.  We found that while the results from the two calculations differ, the experimental signature in both cases is the same and equivalent to that of a thermal ion in a static trap.

We thank Dave Kielpinski for invaluable help with the experimental details.  We also thank Paul Alsing, Bill Unruh, John Preskill, Jeff Kimble, Greg Ver Steeg, and Michael Nielsen for useful discussions and suggestions.  NCM extends much appreciation to the faculty and staff of the Caltech Institute for Quantum Information for their hospitality during his visit, which helped bring this work to fruition.  NCM was supported by the United States Department of Defense, and GJM acknowledges support from the Australian Research Council.


\bibliography{main}

\begin{thebibliography}{19}
\expandafter\ifx\csname natexlab\endcsname\relax\def\natexlab#1{#1}\fi
\expandafter\ifx\csname bibnamefont\endcsname\relax
  \def\bibnamefont#1{#1}\fi
\expandafter\ifx\csname bibfnamefont\endcsname\relax
  \def\bibfnamefont#1{#1}\fi
\expandafter\ifx\csname citenamefont\endcsname\relax
  \def\citenamefont#1{#1}\fi
\expandafter\ifx\csname url\endcsname\relax
  \def\url#1{\texttt{#1}}\fi
\expandafter\ifx\csname urlprefix\endcsname\relax\def\urlprefix{URL }\fi
\providecommand{\bibinfo}[2]{#2}
\providecommand{\eprint}[2][]{\url{#2}}

\bibitem[{\citenamefont{Stoler}(1970)}]{Stoler1970}
\bibinfo{author}{\bibfnamefont{D.}~\bibnamefont{Stoler}},
  \bibinfo{journal}{Phys. Rev. D} \textbf{\bibinfo{volume}{1}},
  \bibinfo{pages}{3217} (\bibinfo{year}{1970}).

\bibitem[{\citenamefont{Yuen}(1976)}]{Yuen1976}
\bibinfo{author}{\bibfnamefont{H.~P.} \bibnamefont{Yuen}},
  \bibinfo{journal}{Phys. Rev. A} \textbf{\bibinfo{volume}{13}},
  \bibinfo{pages}{2226} (\bibinfo{year}{1976}).

\bibitem[{\citenamefont{Hollenhorst}(1979)}]{Hollenhorst1979}
\bibinfo{author}{\bibfnamefont{J.~N.} \bibnamefont{Hollenhorst}},
  \bibinfo{journal}{Phys. Rev. D} \textbf{\bibinfo{volume}{19}},
  \bibinfo{pages}{1669} (\bibinfo{year}{1979}).

\bibitem[{\citenamefont{Ma and Rhodes}(1989)}]{Ma1989}
\bibinfo{author}{\bibfnamefont{X.}~\bibnamefont{Ma}} \bibnamefont{and}
  \bibinfo{author}{\bibfnamefont{W.}~\bibnamefont{Rhodes}},
  \bibinfo{journal}{Phys. Rev. A} \textbf{\bibinfo{volume}{39}},
  \bibinfo{pages}{1941} (\bibinfo{year}{1989}).

\bibitem[{\citenamefont{Dodonov and Klimov}(1996)}]{Dodonov1996}
\bibinfo{author}{\bibfnamefont{V.~V.} \bibnamefont{Dodonov}} \bibnamefont{and}
  \bibinfo{author}{\bibfnamefont{A.~B.} \bibnamefont{Klimov}},
  \bibinfo{journal}{Phys. Rev. A} \textbf{\bibinfo{volume}{53}},
  \bibinfo{pages}{2664} (\bibinfo{year}{1996}).

\bibitem[{\citenamefont{Moore}(1970)}]{Moore1970}
\bibinfo{author}{\bibfnamefont{G.~T.} \bibnamefont{Moore}},
  \bibinfo{journal}{J. Math. Phys.}
  \textbf{\bibinfo{volume}{11}}, \bibinfo{pages}{2679} (\bibinfo{year}{1970}).

\bibitem[{\citenamefont{Kim and Kim}(1995)}]{Kim1995}
\bibinfo{author}{\bibfnamefont{S.~P.} \bibnamefont{Kim}} \bibnamefont{and}
  \bibinfo{author}{\bibfnamefont{S.-W.} \bibnamefont{Kim}},
  \bibinfo{journal}{Phys. Rev. D} \textbf{\bibinfo{volume}{51}},
  \bibinfo{pages}{4254} (\bibinfo{year}{1995}).

\bibitem[{\citenamefont{Polarski and Starobinsky}(1996)}]{Polarski1996}
\bibinfo{author}{\bibfnamefont{D.}~\bibnamefont{Polarski}} \bibnamefont{and}
  \bibinfo{author}{\bibfnamefont{A.~A.} \bibnamefont{Starobinsky}},
  \bibinfo{journal}{Classical and Quantum Gravity}
  \textbf{\bibinfo{volume}{13}}, \bibinfo{pages}{377} (\bibinfo{year}{1996}).

\bibitem[{\citenamefont{Kiefer et~al.}(1998)\citenamefont{Kiefer, Lesgourgues,
  Polarski, and Starobinsky}}]{Kiefer1998}
\bibinfo{author}{\bibfnamefont{C.}~\bibnamefont{Kiefer}},
  \bibinfo{author}{\bibfnamefont{J.}~\bibnamefont{Lesgourgues}},
  \bibinfo{author}{\bibfnamefont{D.}~\bibnamefont{Polarski}}, \bibnamefont{and}
  \bibinfo{author}{\bibfnamefont{A.~A.} \bibnamefont{Starobinsky}},
  \bibinfo{journal}{Classical and Quantum Gravity}
  \textbf{\bibinfo{volume}{15}}, \bibinfo{pages}{L67} (\bibinfo{year}{1998}).

\bibitem[{\citenamefont{Leibfried
  et~al.}(2003{\natexlab{a}})\citenamefont{Leibfried, Blatt, Monroe, and
  Wineland}}]{Leibfried2003}
\bibinfo{author}{\bibfnamefont{D.}~\bibnamefont{Leibfried}},
  \bibinfo{author}{\bibfnamefont{R.}~\bibnamefont{Blatt}},
  \bibinfo{author}{\bibfnamefont{C.}~\bibnamefont{Monroe}}, \bibnamefont{and}
  \bibinfo{author}{\bibfnamefont{D.}~\bibnamefont{Wineland}},
  \bibinfo{journal}{Rev. Mod. Phys.} \textbf{\bibinfo{volume}{75}},
  \bibinfo{pages}{281} (\bibinfo{year}{2003}{\natexlab{a}}).

\bibitem[{\citenamefont{Monroe et~al.}(1995)\citenamefont{Monroe, Meekhof,
  King, Jefferts, Itano, Wineland, and Gould}}]{Monroe1995}
\bibinfo{author}{\bibfnamefont{C.}~\bibnamefont{Monroe}},
  \bibinfo{author}{\bibfnamefont{D.~M.} \bibnamefont{Meekhof}},
  \bibinfo{author}{\bibfnamefont{B.~E.} \bibnamefont{King}},
  \bibinfo{author}{\bibfnamefont{S.~R.} \bibnamefont{Jefferts}},
  \bibinfo{author}{\bibfnamefont{W.~M.} \bibnamefont{Itano}},
  \bibinfo{author}{\bibfnamefont{D.~J.} \bibnamefont{Wineland}},
  \bibnamefont{and} \bibinfo{author}{\bibfnamefont{P.~L.} \bibnamefont{Gould}},
  \bibinfo{journal}{Phys. Rev. Lett.} \textbf{\bibinfo{volume}{75}},
  \bibinfo{pages}{4011} (\bibinfo{year}{1995}).

\bibitem[{\citenamefont{James}(1998)}]{James1998}
\bibinfo{author}{\bibfnamefont{D.~F.~V.} \bibnamefont{James}},
  \bibinfo{journal}{Applied Physics B: Lasers and Optics}
  \textbf{\bibinfo{volume}{66}}, \bibinfo{pages}{181} (\bibinfo{year}{1998}).

\bibitem[{\citenamefont{Leibfried
  et~al.}(2003{\natexlab{b}})\citenamefont{Leibfried, \surname{De Marco},
  Meyer, Lucas, Barrett, Britton, Itano, Jelenkovic, Langer, Rosenband
  et~al.}}]{Leibfried2003a}
\bibinfo{author}{\bibfnamefont{D.}~\bibnamefont{Leibfried}},
  \bibinfo{author}{\bibfnamefont{B.}~\bibnamefont{\surname{De Marco}}},
  \bibinfo{author}{\bibfnamefont{V.}~\bibnamefont{Meyer}},
  \bibinfo{author}{\bibfnamefont{D.}~\bibnamefont{Lucas}},
  \bibinfo{author}{\bibfnamefont{M.}~\bibnamefont{Barrett}},
  \bibinfo{author}{\bibfnamefont{J.}~\bibnamefont{Britton}},
  \bibinfo{author}{\bibfnamefont{W.~M.} \bibnamefont{Itano}},
  \bibinfo{author}{\bibfnamefont{B.}~\bibnamefont{Jelenkovic}},
  \bibinfo{author}{\bibfnamefont{C.}~\bibnamefont{Langer}},
  \bibinfo{author}{\bibfnamefont{T.}~\bibnamefont{Rosenband}},
  \bibnamefont{et~al.}, \bibinfo{journal}{Nature}
  \textbf{\bibinfo{volume}{422}}, \bibinfo{pages}{412}
  (\bibinfo{year}{2003}{\natexlab{b}}).

\bibitem[{\citenamefont{Schmidt-Kaler et~al.}(2003)\citenamefont{Schmidt-Kaler,
  H\"affner, Riebe, Lancaster, Deuschle, Becher, Roos, Eschner, and
  Blatt}}]{Schmidt-Kaler2003}
\bibinfo{author}{\bibfnamefont{F.}~\bibnamefont{Schmidt-Kaler}},
  \bibinfo{author}{\bibfnamefont{H.}~\bibnamefont{H\"affner}},
  \bibinfo{author}{\bibfnamefont{M.}~\bibnamefont{Riebe}},
  \bibinfo{author}{\bibfnamefont{G.~P.~T.} \bibnamefont{Lancaster}},
  \bibinfo{author}{\bibfnamefont{T.}~\bibnamefont{Deuschle}},
  \bibinfo{author}{\bibfnamefont{C.}~\bibnamefont{Becher}},
  \bibinfo{author}{\bibfnamefont{C.~F.} \bibnamefont{Roos}},
  \bibinfo{author}{\bibfnamefont{J.}~\bibnamefont{Eschner}}, \bibnamefont{and}
  \bibinfo{author}{\bibfnamefont{R.}~\bibnamefont{Blatt}},
  \bibinfo{journal}{Nature} \textbf{\bibinfo{volume}{422}},
  \bibinfo{pages}{408} (\bibinfo{year}{2003}).

\bibitem[{\citenamefont{Alsing et~al.}(2005)\citenamefont{Alsing, Dowling, and
  Milburn}}]{Alsing2005}
\bibinfo{author}{\bibfnamefont{P.~M.} \bibnamefont{Alsing}},
  \bibinfo{author}{\bibfnamefont{J.~P.} \bibnamefont{Dowling}},
  \bibnamefont{and} \bibinfo{author}{\bibfnamefont{G.~J.}
  \bibnamefont{Milburn}}, \bibinfo{journal}{Phys. Rev. Lett.}
  \textbf{\bibinfo{volume}{94}}, \bibinfo{eid}{220401} (\bibinfo{year}{2005}).

\bibitem[{\citenamefont{Gibbons and Hawking}(1977)}]{Gibbons1977}
\bibinfo{author}{\bibfnamefont{G.~W.} \bibnamefont{Gibbons}} \bibnamefont{and}
  \bibinfo{author}{\bibfnamefont{S.~W.} \bibnamefont{Hawking}},
  \bibinfo{journal}{Phys. Rev. D} \textbf{\bibinfo{volume}{15}},
  \bibinfo{pages}{2738} (\bibinfo{year}{1977}).

\bibitem[{\citenamefont{Wallentowitz and Vogel}(1996)}]{Wallentowitz1996}
\bibinfo{author}{\bibfnamefont{S.}~\bibnamefont{Wallentowitz}}
  \bibnamefont{and} \bibinfo{author}{\bibfnamefont{W.}~\bibnamefont{Vogel}},
  \bibinfo{journal}{Phys. Rev. A} \textbf{\bibinfo{volume}{54}},
  \bibinfo{pages}{3322} (\bibinfo{year}{1996}).

\bibitem[{\citenamefont{Carvalho et~al.}(2004)\citenamefont{Carvalho, Furtado,
  and Pedrosa}}]{Carvalho2004}
\bibinfo{author}{\bibfnamefont{A.~M.~de~M.}~\bibnamefont{Carvalho}},
  \bibinfo{author}{\bibfnamefont{C.}~\bibnamefont{Furtado}}, \bibnamefont{and}
  \bibinfo{author}{\bibfnamefont{I.~A.} \bibnamefont{Pedrosa}},
  \bibinfo{journal}{Phys. Rev. D} \textbf{\bibinfo{volume}{70}}, \bibinfo{eid}{123523}
  (pages~\bibinfo{numpages}{6}) (\bibinfo{year}{2004}).

\bibitem[{\citenamefont{Unruh}(1981)}]{Unruh1981}
\bibinfo{author}{\bibfnamefont{W.~G.} \bibnamefont{Unruh}},
  \bibinfo{journal}{Phys. Rev. Lett.} \textbf{\bibinfo{volume}{46}},
  \bibinfo{pages}{1351} (\bibinfo{year}{1981}).

\end{thebibliography}

\end{document}